\documentclass[authoryear]{FLO_v1}%

\usepackage{graphicx}
\usepackage{upgreek}
\usepackage{multicol,multirow}
\usepackage{amsmath,amssymb,amsfonts}
\usepackage{mathrsfs}
\usepackage{amsthm}
\usepackage[figuresright]{rotating}
\usepackage{appendix}
\usepackage[authoryear]{natbib}
\usepackage{ifpdf}
\usepackage[T1]{fontenc}
\usepackage{newtxtext}
\usepackage{newtxmath}
\usepackage{textcomp}
\usepackage{xcolor}

\usepackage[colorlinks,allcolors=blue]{hyperref}
\definecolor{jourcolor}{cmyk}{1,0.57,0.01,0.38}
\hypersetup{
    colorlinks,%
    citecolor=jourcolor,%
    filecolor=jourcolor,%
    linkcolor=jourcolor,%
    urlcolor=jourcolor
}

\theoremstyle{definition}

\articletype{RESEARCH ARTICLE}

\DOI{}

\Year{}

\Vol{}

\Price{}

\art-id{FLO2000049}

\citearticle{Goldshmid, R.H., \& Dabiri, J.O.}

\begin{document}
\title[Physical constraints on visual anemometry]{Physical constraints on visual anemometry using vegetation displacement statistics}

\author [Roni H. Goldshmid and John O. Dabiri]{Roni H. Goldshmid$^{1}$ {\href{https://orcid.org/0000-0001-9095-3259}{\includegraphics[width=0.02\textwidth]{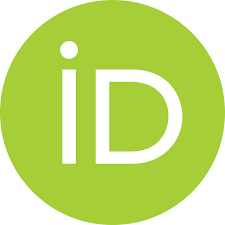}}}}

\author[Roni H. Goldshmid and John O. Dabiri]{John O. Dabiri$^{1^\ast,2}$ {\href{https://orcid.org/0000-0002-6722-9008}{\includegraphics[width=0.02\textwidth]{orcid_logo}}}}
 
\address[1]{Graduate Aerospace Laboratories, California Institute of Technology, Pasadena, CA 91125, USA}
\address[2]{Mechanical Engineering, California Institute of Technology, Pasadena, CA 91125, USA}

\corres{*}{Corresponding author. E-mail:
\emaillink{jodabiri@caltech.edu}}


\date{\textbf{Received:} XX 2022; \textbf{Revised:} XX XX 2022; \textbf{Accepted:} XX XX 2022}

\abstract{Visual anemometry (VA) leverages observations of fluid-structure interactions to infer incident flow characteristics. Recent work has demonstrated the concept of VA using both data-driven and physical modelling approaches applied to natural vegetation. These methods have not yet achieved generalization across plant species and require site-specific calibration. We conducted a laboratory study in an open circuit wind tunnel using overhead imagery of three vegetation species to assess the utility of vegetation displacement fields for wind speed inference. Both the wind and vegetation speeds exhibited a two-parameter Weibull distribution. The relationship between the scale factor (one of the two parameters) of the wind and vegetation was found to be well described by a sigmoid function, indicating three regions of distinct structural response to the wind loading at low, intermediate, and high wind speeds. Within an intermediate range of wind speeds, the wind and vegetation scale factors are nearly linearly proportional, thereby facilitating VA. Importantly, the wind and vegetation scale factors were found to be uncorrelated in low and high wind regimes, revealing a fundamental constraint on VA using structure response data. We discuss the physical basis for these regimes and present additional parametric relationships that can be exploited in the intermediate wind regime to potentially generalize inference of the wind speed and direction.}

\maketitle

\begin{boxtext}

\textbf{\mathversion{bold}Impact Statement}
We assess the feasibility of visual anemometry (VA), i.e., inference of incident wind based on observation of structural response, using overhead observations and several species of vegetation. We discover a fundamental constraint of VA when the vegetation kinematics are utilized for estimations, and we discuss its physical interpretation. Additionally, we propose two methods to infer the wind speed and direction in the range of flow conditions suitable for VA. These methods can be used to leverage environmental flow-structure interactions in applications including weather prediction, pollution tracking, aviation safety monitoring, and wildland firefighting.  
\end{boxtext}

\section{Introduction}
\label{sec:intro}
Visual anemometry (VA) refers to a class of measurement techniques wherein observations of fluid-structure interactions are used to infer incident flow characteristics. Early efforts toward VA include the Beaufort scale \citep{Kinsman1969,Wade1979} in which visual cues are used to qualitatively estimate mean flow speeds on land and at sea, e.g., using the rising angles of smoke plumes and whitecap formation on the sea surface, respectively. Recent quantitative predictions of fluid dynamic quantities (e.g., water wave amplitude, height, and sea state) were acquired using machine learning applied to water wave videos from oceans and lakes \citep{Spencer2004}. Another study recently demonstrated the prediction of incident wind velocity and intrinsic cloth material properties of a flapping flag \citep{Runia2019}. 

Further proof-of-concept of VA was presented by \citet{Cardona2019}, where a data-driven approach was applied to natural vegetation. This study demonstrated the ability to infer average wind speed from video recordings of a tree using neural networks. While the model successfully interpolated among wind conditions within the training set, with prediction errors approaching the background turbulence intensity, inaccuracies at low and high winds were revealed \citep{Cardona2019}. Physical interpretation of the data-driven approach, including explanation of the reduced performance at low and high wind speeds, has been limited by the lack of direct reference to the underlying flow physics. 

The interactions of trees with the wind have been widely studied \citep{deLangre2008}, and recent attempts at physics-informed VA \citep{Cardona2021,Cardona2022} focused on the dynamics of wind interactions with natural vegetation. \citet{Cardona2021} explored the correlation of the mean deflection of trees and other cantilevered structures with the mean wind speeds and found agreement between VA estimates and reference wind speeds. With the notable exception of accuracy at low wind speeds, \citet{Cardona2021} found good agreement between the VA-estimated wind speed and the ground truth. Subsequently, \citep{Cardona2022} developed a physical model relating the average wind to the amplitude of oscillations of vegetation. This method was proven to be more robust than VA based on mean deflection. However, in both this study and in companion work \citep{Sun2022}, errors persisted in low and high wind speed regimes. In sum, the findings of both physics-based models exhibited notable errors at low and high wind speeds.

The present work specifically explores the origin of persistent inaccuracies at low and high winds. We conducted a suite of measurements in an open circuit wind tunnel to study VA using overhead imagery of three species of vegetation. Overhead views were selected to assess the utility of the measured vegetation displacement fields for wind speed inference. We found that both wind speeds and vegetation speeds exhibited a two-parameter Weibull distribution. The relationship between the scale factor (one of the two Weibull parameters) of the wind and vegetation was observed to follow a sigmoid function, indicating three regions of distinct structural response to the wind loading at low, intermediate, and high wind speeds. The intermediate range of wind speeds was found to be most suitable for VA because it is the region where the wind scale factor is nearly linearly proportional to the vegetation motion scale factor. Importantly, the wind and vegetation scale factors were found to be uncorrelated in low and high wind regimes, revealing a fundamental constraint on the use of vegetation displacement statistics for VA outside of intermediate wind speeds. Additional parametric relationships within the regime of linear proportionality were utilized to infer the wind speed and direction across different vegetation species. 

\section {Experimental Methods}
\subsection{Setup}
\begin{figure}[!htb]
      \centerline{\includegraphics[width=.95\textwidth]{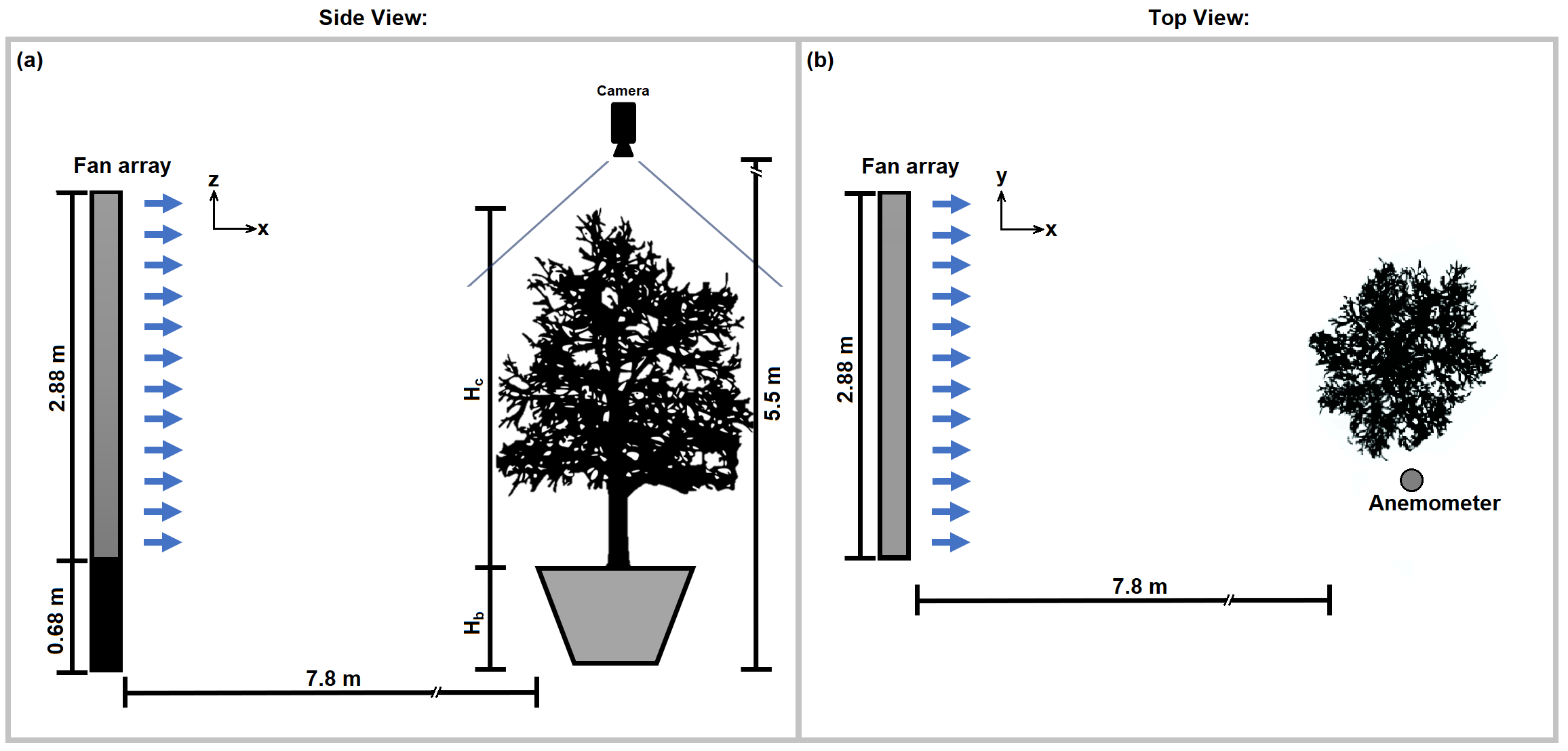}}
      \caption{(a) Side view of the experimental setup, including the fan array, example vegetation, and overhead camera (vertical distance and camera size not drawn to scale). $H_c$ and $H_b$ of each vegetation are detailed in table \ref{tab:treedims}. (b) Top view of the setup, including the fan array, example vegetation, and anemometer.}
    \label{fig:1}
    \end{figure}

Experiments were conducted in an open-circuit wind tunnel at the Caltech Center for Autonomous Systems and Technologies. The schematic of the experimental setup is shown in figure \ref{fig:1}. The frontal area of the fan array ($A_f$) is 2.88 m × 2.88 m. It is elevated from the ground by a base height ($H_{f_b}$) of 0.68 m. The selected incident wind speed range $(U_0)$ was from 1 to 8 m/s and the turbulence intensity ranged from 19\% to 56\% at 7.8 meters downstream from the fan array, where each measured vegetation was located. The relatively large turbulence intensities are associated with the open-circuit wind tunnel flow behavior at the downstream distances used for these experiments. These values are consistent with field studies of VA under naturally occurring, near-ground, wind conditions \citep{Cardona2019,Cardona2022}.

Three species of vegetation were examined: a coast oak tree (\textit{Quercus agrifolia}), a camphor tree (\textit{Cinnamomum camphora}), and a patch of bunchgrasses comprised of twenty clusters of bullgrass (\textit{Muhlenbergia emersleyi}). Each species was described by a distinct ratio between its canopy height ($H_c$) and trunk diameter ($D_c$), also known as the slenderness ($S$) ratio, i.e. $S = H_c/D_c$. Table \ref{tab:treedims} details the slenderness measures for each of the examined vegetation and the corresponding $H_c$ and $D_c$. Each of the species was examined under the same incident wind speeds. A second set of tests was conducted for each of the tree species that included the bullgrass patch 1.22 m upstream of the tree. No significant variations were observed between the tree experiments with and without the bullgrass insert present upstream; hence, these tests were combined for subsequent analyses. The maximum blockage ratio, $A/A_f$, where $A$ is approximate frontal area of the vegetation under no load, ranged from 7.5\% to 57\%. The resulting range of Reynolds numbers was $O(10^4-10^6)$, based on a length scale of $A^{1/2}$. All tree specific measures are summarized in table \ref{tab:treedims}.

 \begin{table}[!htb]
      \begin{center}
    \def~{\hphantom{0}}
      \begin{tabular}{lcccccccc}
                      &$A,$ m$^2$   & $\frac{A}{A_f}$   &$D_c$, cm  & $\frac{H_b}{H_{b_f}}$ & $H_c$, m  & Re$_{\sqrt{A}}\times10^{-6}$ & $S$       & TI, \% \\[3pt]
           Oak        &~4.74 ~      & ~0.571 ~          &~8.46~     & ~0.86~                & ~2.77 ~   & 0.20 -- 1.27                  & ~0.031 ~   & 21 -- 29\\    
           Camphor    &~4.70 ~      & ~0.567 ~          &~3.85~     & ~0.86~                & ~2.57 ~   & 0.34 -- 1.24                  & ~0.015 ~   & 19 -- 25\\
           Bullgrass  &~0.622~      & ~0.075~           &~0.18~     & ~0.62~                & ~0.51 ~   & 0.06 -- 0.44                  & ~0.0035~   & 27 -- 56\\
      \end{tabular}
      \caption{Vegetation specific parameters describing each species and wind conditions studied. $A$ and $A_f$ are vegetation and fan array frontal areas, respectively, $D_c$ is trunk diameter, $H_b$ and $H_{b_f}$ are vegetation and fan array base height, respectively, $H_c$ is canopy height, Re is Reynolds number, $S$ is slenderness, and TI is turbulence intensity.}
      \label{tab:treedims}
      \end{center}
    \end{table}

Wind measurements were collected using two anemometers and visual measurements were collected using an overhead camera. The anemometers were placed at 1.8 m and 2.54 m above the ground, respectively. The difference in mean wind speeds measured by the two anemometers was less than the turbulence intensity, and the data from the anemometer at 2.54 m was used for this study. The anemometer measured the streamwise ($x$) and transverse ($y$) wind velocity components at a 4 Hz sampling rate. A color CMOS camera (Campbell Scientific CCFC with image sensor Omnivision OV5653) was installed 5.5 m above the ground, or 175\% of the height of the tallest tree, to measure the vegetation from overhead. The camera recorded videos at 15 frames per second with a resolution of 1280 pixels x 720 pixels. Example overhead views at various wind speeds are presented in figure \ref{fig:2}. The camera and anemometer were simultaneously triggered by a data logger (Campbell Scientific CR1000X) and each recorded 58-second samples of the wind-vegetation interactions. Thus, each sample consists of 870 frames and 232 two-component (u,v) anemometry measurements. This duration was sufficient to capture variability in the incident wind speed within each sample. 

\subsection{Analysis}
\begin{figure}[!htb]
      \centerline{\includegraphics[width=\textwidth]{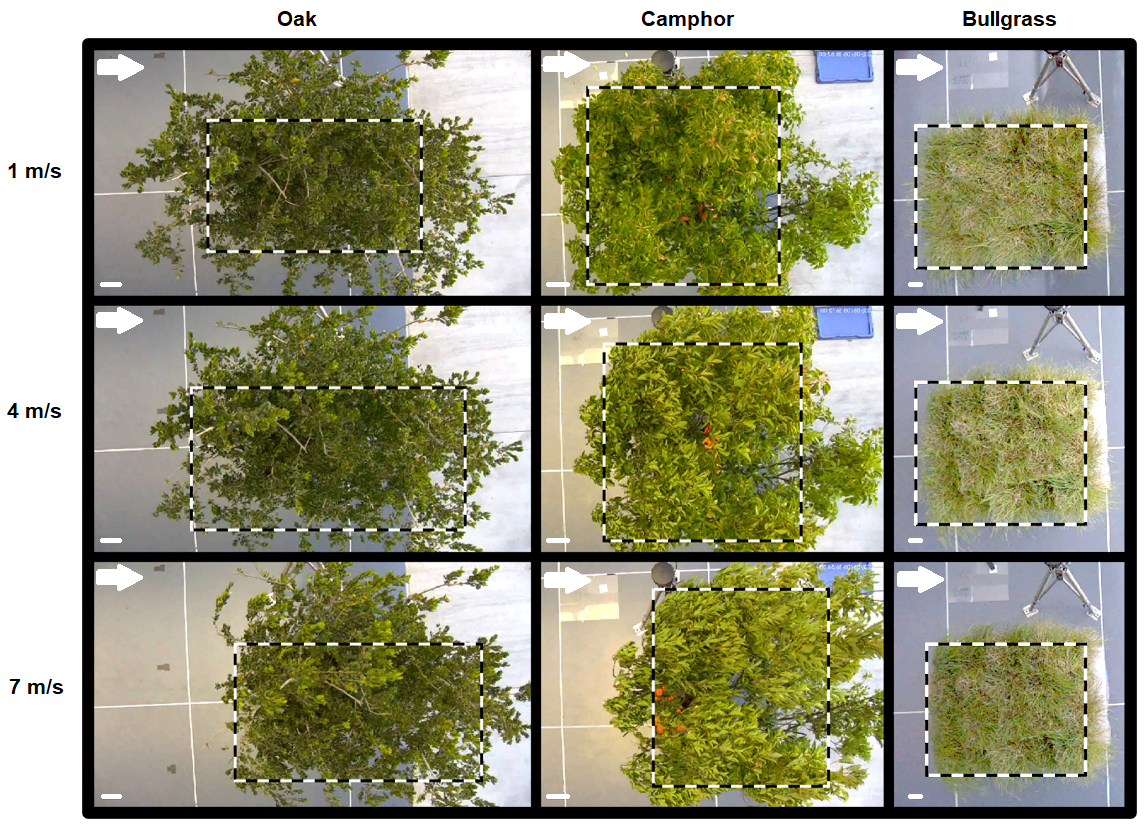}}
      \caption{Example overhead views of the vegetation response to three incident wind speeds. The dashed black and white rectangles represent the bounding boxes. The incident wind direction is presented using the white arrows pointing from left to right in each frame. A reference length scale of 0.1 m is marked at the southwest corner of each tile.}
    \label{fig:2}
    \end{figure}

Vegetation displacement fields were computed using PIVlab \citep{PIVlab}, a Matlab based toolbox for particle image velocimetry (PIV), after converting each frame to a monochrome image. An interrogation window size of 16 x 16 pixels with a 50\% overlap was used throughout. We manually selected bounding boxes for each sample to only contain the canopy and exclude the tunnel floor in the background. Sample bounding boxes are displayed as dashed black and white rectangles in figure \ref{fig:2}. PIV spatial calibration was based on a reference length scale at the height of the top of each canopy. The data were grouped by vegetation species and binned into 1 m/s bins using the wind speed measured by the anemometer located at 2.54 m above the ground. 

Temporal statistics of the incident wind are commonly described using two-parameter Weibull distributions for the probability density \citep{Bowden1983,Tuller1984},
    \begin{equation}
     f(x)=\begin{cases}
              \cfrac{k}{\lambda}\left(\cfrac{x}{\lambda}\right)^{k-1} \quad \quad x\geq0 \quad,\\
              0 \ \   \quad  \quad \quad \quad \quad x<0 \quad.\\
         \end{cases}
      \label{Weib}
    \end{equation}
The unbiased maximum likelihood estimator (MLE) of the two parameters, namely, the scale factor $(\lambda)$ and the shape factor $(k)$, are the solution of the simultaneous equations \citep{Bowden1983},
    \begin{equation}
    \lambda=\Bigg[ \Big(\cfrac{1}{n}\Big) \sum_{i=1}^{n} x_i^{k} \Bigg]^{\cfrac{1}{k}} \quad,
      \label{lam_mle}
    \end{equation}
    and,
    \begin{equation}
        k=\cfrac{n}{\frac{1}{\lambda}\sum_{i=1}^{n} x_i^{k}\log{x_i}-\sum_{i=1}^{n} \log{x_i}} \quad,
      \label{k_mle}
    \end{equation}
where $n$ is the number of samples and $x$ is an observation. We used these equations to describe each sample comprising anemometer measurements of the wind speed and also examined the relevance of this distribution for the vegetation displacement statistics. To assess the fit of the measurements to the Weibull distribution, we computed quantile-quantile (q-q) analyses \citep{Wilk1968}. Quantiles are defined as continuous intervals resulting from division of the full range of a probability distribution into equal probabilities by ordering the sample data from smallest to largest. Each measurement quantile is compared to the corresponding expected theoretical value. A resulting linear relationship with the theoretical distribution suggests that sampled data likely comes from that distribution. 

\section{Results}
\subsection{Distribution of Wind and Vegetation Speeds}
Example probability density functions (PDF) of the wind speed measured in the presence of the vegetation and their corresponding Weibull distribution fits are presented in figure \ref{fig:3}(a)-(c). The fit parameters and confidence intervals are summarized in table S1 in the supplementary material. Figure \ref{fig:3}(d)-(f) present the q-q curves of the intermediate wind speed distributions. The corresponding low and high wind examples can be seen in figures S1-S3 of the supplementary material. The solid reference lines illustrate the region that corresponds to 75\% of the observations, and the resulting linear correlation in the majority of the data suggests the sample data likely comes from the Weibull distribution. The deviation of the data above/below the reference line is only observed in the tails indicating more/less occurrence in the tails than predicted by the theoretical distribution. 

\begin{figure}[!htb]
      \centerline{\includegraphics[width=\textwidth]{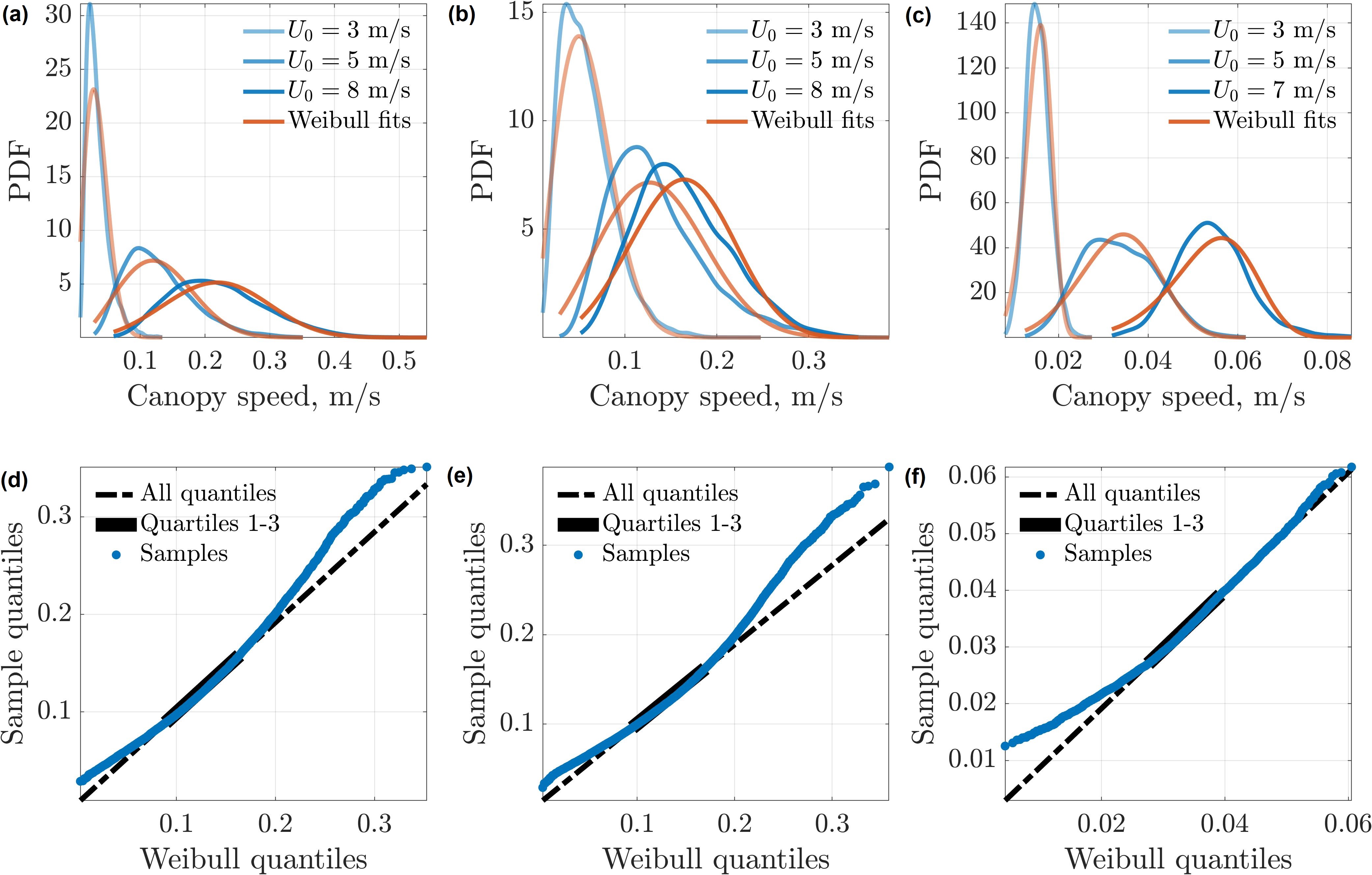}}
      \caption{The probability distributions of the vegetation displacement speeds of the (a) oak, (b) camphor, and (c) bullgrass are shown in blue. Their corresponding Weibull distribution fits are shown in orange. The MLE estimates and confidence intervals of the fits can be found in table S1 of the supplementary material. (d)-(f) Example q-q analyses of the intermediate wind speed observed in (a)-(c), respectively. The solid reference line connects the first and third quartiles, corresponding to 75\% of the data, and the dashed reference line extends the solid line to the ends of the samples.}
    \label{fig:4}
    \end{figure}

Since the wind is the only dynamic load on the vegetation, we compared the distributions of the vegetation displacement speeds with the distributions of the wind speeds. For a more direct comparison with the pointwise wind measurements, the vegetation displacement fields were spatially averaged in each frame. Figure \ref{fig:4} presents PDF, Weibull fits, and q-q curves of the vegetation speeds. The fit parameters can be found in table S1 and q-q curves of the low and high speeds in figures S1-S3, in the supplementary material. As observed in the wind data, the first three quartiles of the vegetation are also linear, indicating a good agreement with the theoretical distribution. A larger deviation from the Weibull distribution, relative to the wind speed measurements, was only observed in the tails of the vegetation speeds. A tendency towards a more linear shape was observed in the vegetation kinematics with an increase in wind speed, indicating the tails of the vegetation displacement distributions are better described by a Weibull distribution at stronger winds.

    \begin{figure}[!htb]
      \centerline{\includegraphics[width=\textwidth]{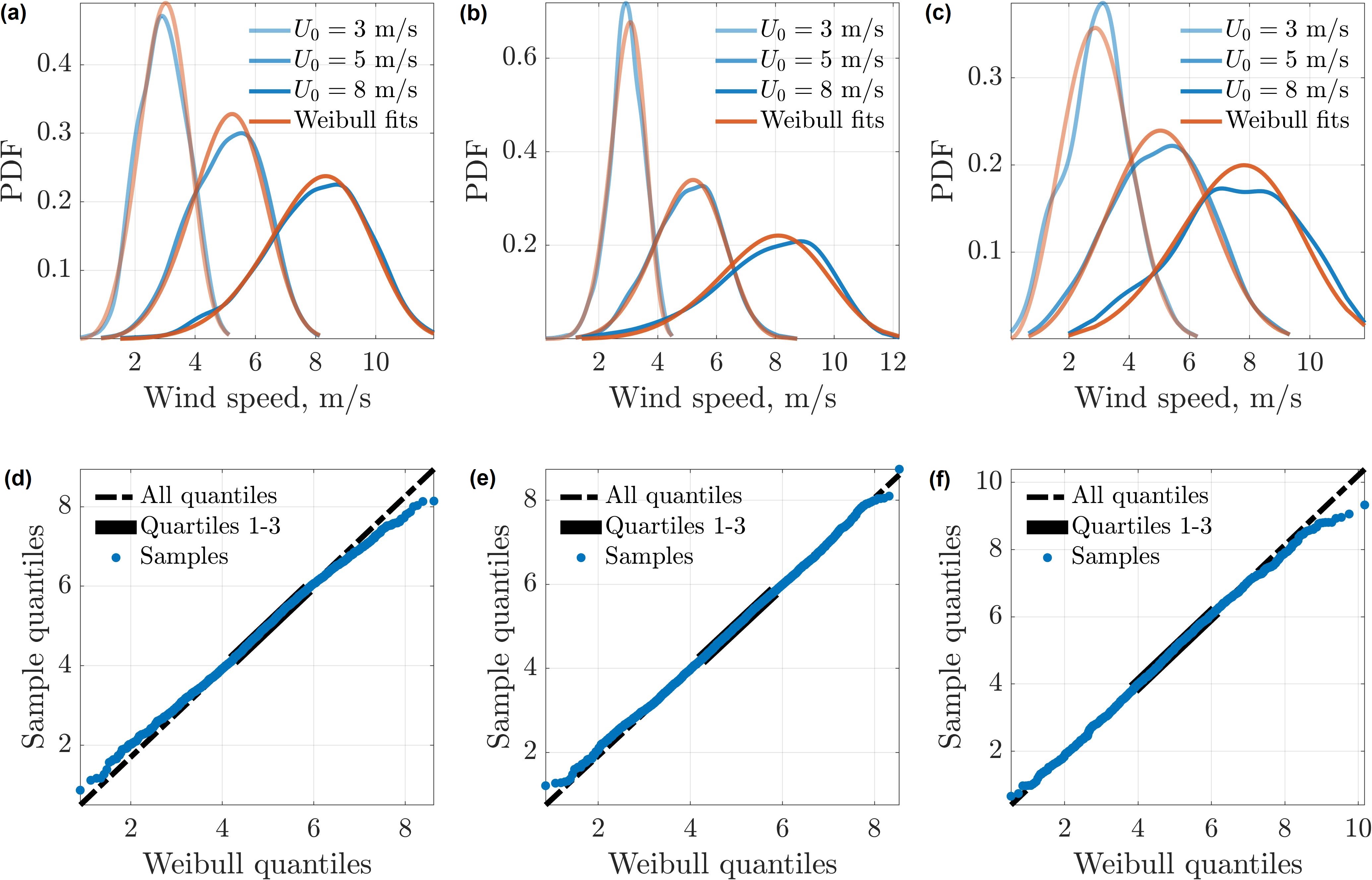}}
      \caption{The probability distributions of the wind measured in the presence of the (a) oak, (b) camphor, and (c) bullgrass are shown in blue. Their corresponding Weibull distribution fits are shown in orange. The MLE estimates and confidence intervals of the fits can be found in table S1 of the supplementary material. (d)-(f) Example q-q analyses of the intermediate wind speed observed in (a)-(c), respectively. The solid reference line connects the first and third quartiles, corresponding to 75\% of the data, and the dashed reference line extends the solid line to the ends of the samples.}
    \label{fig:3}
    \end{figure}

\subsection{Correlation of Wind and Vegetation Speeds} 
\begin{figure}[!htb]
      \centerline{\includegraphics[width=\textwidth]{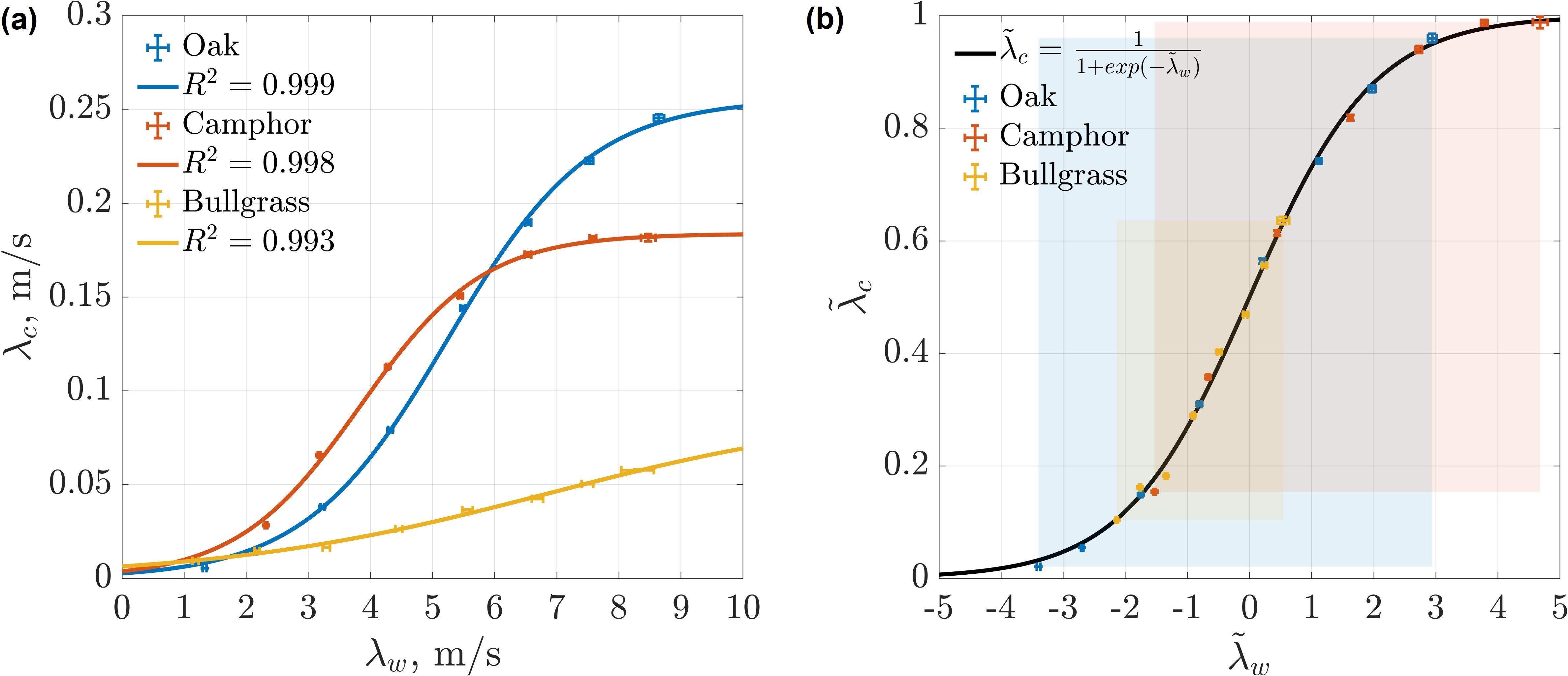}}
      \caption{(a) Relationship between the scale factors of the wind speed and vegetation speed. The error bars represent the confidence intervals of the MLE. These scale factor relationships are fit to three parameter sigmoid functions, and the parameters are presented in table \ref{tab:fitCoefs}. (b) Relationship between the dimensionless scale factors. The translucent rectangles of each vegetation demonstrate the range it occupies within the normalized curve.}
    \label{fig:5}
    \end{figure}

The relationship between the scale factors of the wind speed, $\lambda_w$, and vegetation canopy speed, $\lambda_c$ are quantitatively examined in figure \ref{fig:5}(a). A three parameter sigmoid function fit was computed for each vegetation using,  
    \begin{equation}
     \lambda_c=\cfrac{a}{1+\exp{\bigg( \cfrac{-(\lambda_w-\lambda_{w_0})}{b}\bigg)}} \quad,
      \label{laml}
    \end{equation} 
where $a$ scales the asymptote of the vegetation scale factor in high winds; $\lambda_{w_0}$ determines the center of the region of approximately linear response of the vegetation motion with wind speed; and $b$ determines the range of wind speeds over which the vegetation response is approximately linear. The overall R$^2$ values are presented in the legend and the fit parameters are available in table \ref{tab:fitCoefs}. Using the fit parameters, we define dimensionless coefficients, $\tilde{\lambda}_c=a\lambda_c$ and $\tilde{\lambda}_w=(\lambda_w-\lambda_{w_0})/b$, to collapse the data to a single curve, 
    \begin{equation}
     \tilde{\lambda}_c=\cfrac{1}{1+\exp{(-\tilde{\lambda}_w)}} \quad,
      \label{lam}
    \end{equation}
as presented in figure \ref{fig:5}(b). The sigmoid shape suggests three regimes of vegetation displacement correlation with the flow at low, intermediate, and high wind speeds. The wind and vegetation only exhibit a linear correlation in the intermediate regime, which accordingly limits the use of VA to that regime. It is the region where the scale factor of the wind and vegetation speeds are approximately linearly proportional. In the low- and high-wind regimes, represented by the two plateaus of the sigmoid shape, the vegetation kinematics are insensitive to changes in wind speed. This occurs in the low-wind case because the wind is typically below threshold loading required to displace the vegetation. Higher resilience to wind loading corresponds to vegetation with lower slenderness estimates \citep{Peltola1996}, and we found that the dimensionless ratio of $\lambda_{w_0}/b$ was proportional to the vegetation slenderness estimates (see figure S4 in supplementary material). Conversely, in the high-wind regime, the vegetation becomes maximally deflected, and further increases in wind speed cannot achieve further elastic deformation. Because of these two distinct physical effects, VA cannot be used for accurate wind speed prediction in low- or high-wind regimes. As illustrated in figure \ref{fig:5}(a), the quantitative wind speed threshold between sigmoid regimes varies depending on the specific type of vegetation and further exploitation of the sigmoid fit parameters may predict physical properties of the observed structures. 

Notably, this sigmoid response shape was previously observed in both data driven and physics based approaches \citep{Cardona2019,Cardona2021,Cardona2022,Sun2022}. \citet{Cardona2019} attributed the errors at the low and high wind regimes to the sampling rate and video duration selection, but the present results suggest a more fundamental physical constraint could be relevant. It is possible, however, that other factors of the kinematics may be present concurrently with the physical limitation revealed here. 

   \begin{table}[!htb]
    \begin{center}
    \begin{tabular}{ccccc}
              & $\lambda_{w_0}$, m/s    & $a$, m/s      & $b$, m/s  & $\lambda_{w_0}/{b}$   \\
    Oak       & ~5.250~                 & ~0.256~       & ~1.154~   &  ~4.55~           \\
    Camphor   & ~3.836~                 & ~0.184~       & ~0.991~   &  ~3.87~           \\
    Bullgrass & ~6.862~                 & ~0.091~       & ~2.657~   &  ~2.58~
    \end{tabular}
    \caption{The sigmoid function fit parameters of each vegetation. This table corresponds to figure \ref{fig:5}(a).}
    \label{tab:fitCoefs}
    \end{center}
    \end{table}

\subsection{Inference of Wind Speed and Direction in the Intermediate Wind Regime}
\begin{figure}[!htb]
      \centerline{\includegraphics[width=\textwidth]{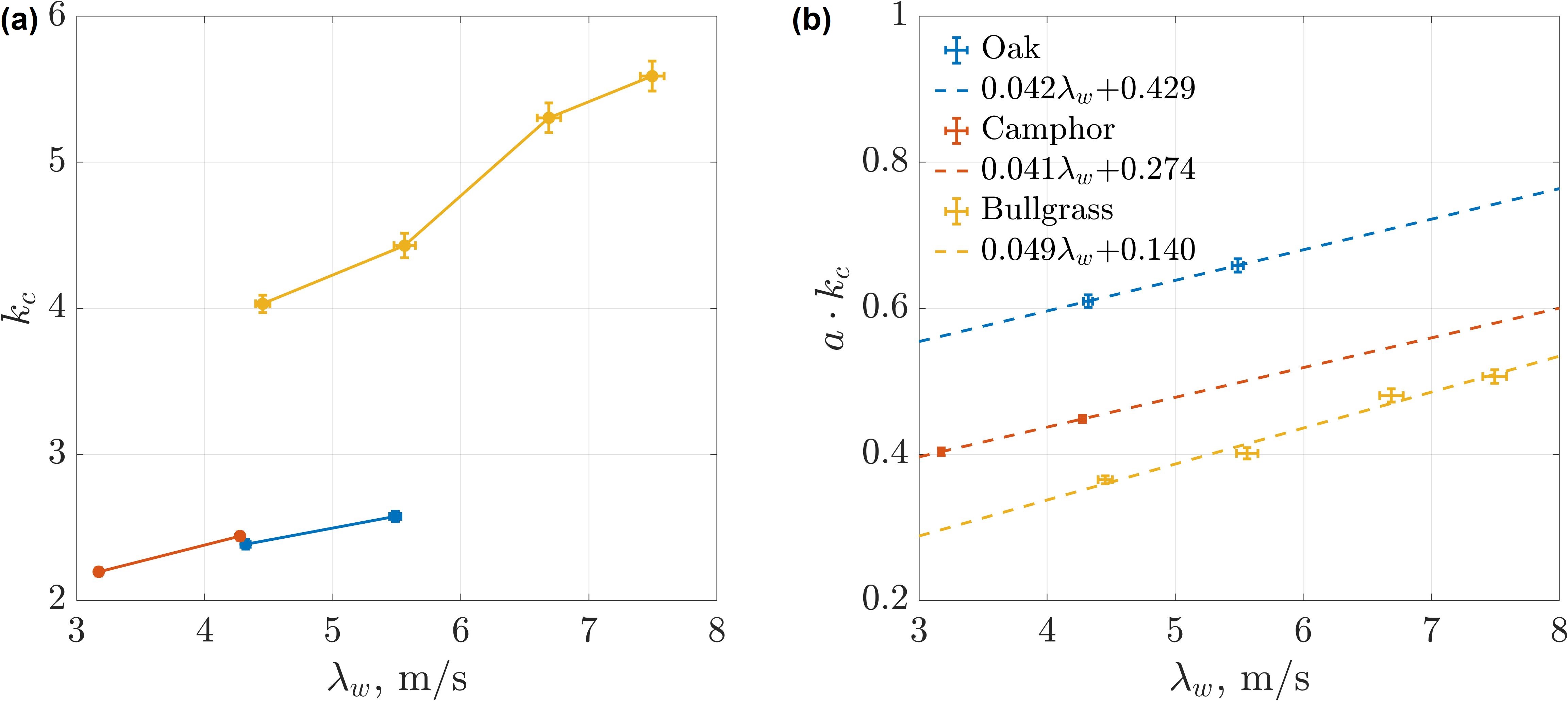}}
      \caption{(a) The relationship between the shape factor of the vegetation displacement fields and the scale factor of the wind within the regime of linear proportionality, i.e. $-1\leq\tilde{\lambda}_w\leq1$.  (b) The relationship between the non-dimensional shape factor of the vegetation displacement fields and the dimensional scale factor of the wind within the same regime of linear proportionality.}
    \label{fig:6}
    \end{figure}

Figure \ref{fig:6}(a) displays the relationship between the shape factor of the vegetation speeds  (i.e. a measure of the distribution breadth) and the scale factor of the wind speeds within the approximately linear regime, i.e. $-1\leq\tilde{\lambda}_w\leq1$. To compare distributions across the three types of vegetation, the shape factor is scaled by $a$ in figure \ref{fig:6}(b). The linear regressions revealed similar slopes of $0.044\pm0.005$ (m/s)$^{-1}$ albeit based on only two available wind speeds within the region of linear correlation in the sigmoid profile. The observed similarity in slope across different vegetation types suggests that measurement of the shape factor of the vegetation displacement distribution as well as as a single corresponding wind speed calibration could be sufficient to achieve site-specific VA. Thus, this approach does not eliminate the calibration requirement discussed in the previous studies \citep{Cardona2019,Cardona2021,Cardona2022,Sun2022}, but is robust to achieve complete automation following an initial site specific calibration.

\begin{figure}[!htb]
      \centerline{\includegraphics[width=\textwidth]{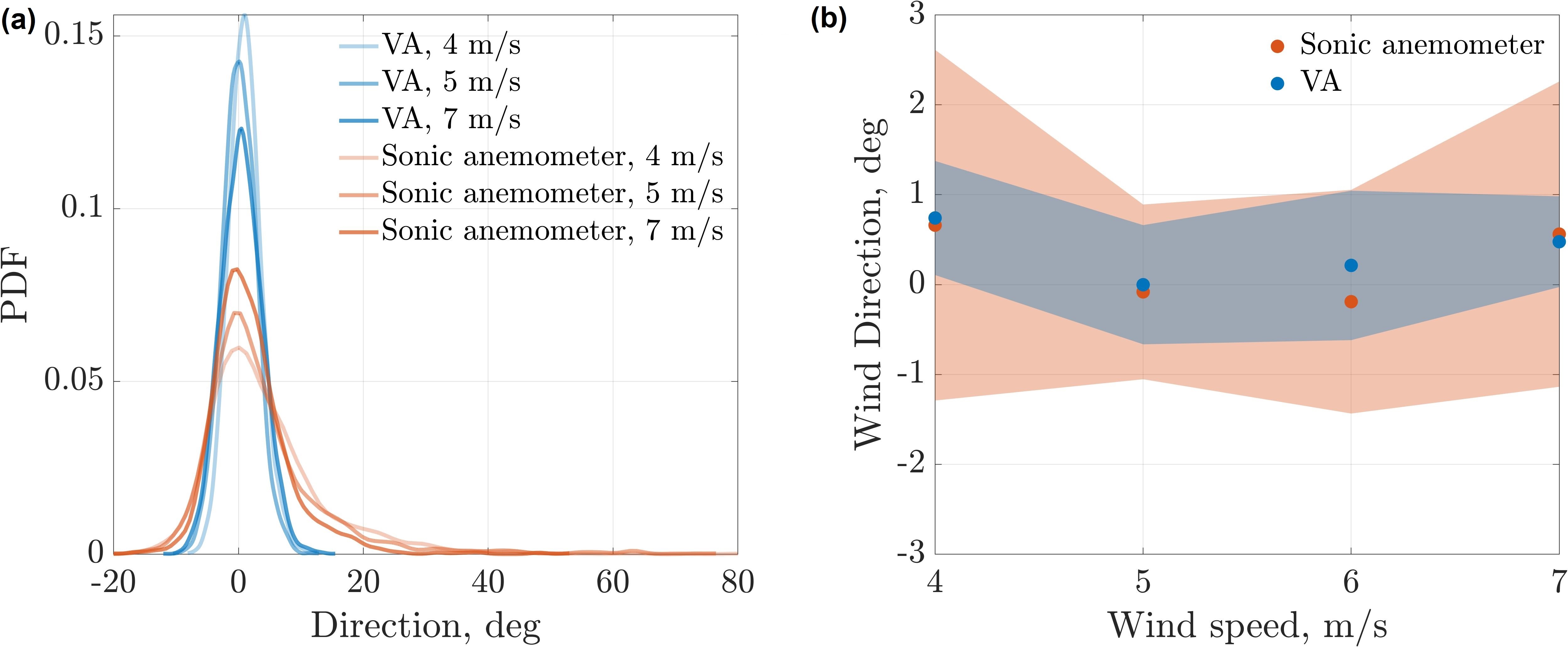}}
      \caption{(a) Probability distributions of the wind direction measured using VA and the sonic anemometer. (b) The binned average (circle markers) and standard deviation (transparent bands) of the predicted wind direction within the linear range of proportionality. The difference in the average direction $-0.12^{\circ}\pm0.24^{\circ}$.}
    \label{fig:7}
    \end{figure}

Finally, we propose an approach to infer the direction of the wind. This approach eliminates the aforementioned calibration requirement. Wind direction estimation via VA exploits the fact that the ratio of wind speed components $(u,v)$ can be inferred even if the absolute magnitude of $u$ and $v$ are unknown individually. Here, we compute the average vegetation displacement direction, i.e. $\theta = tan^{-1} (v/u)$, for each frame in a given observation sample. This analysis is conducted on the bullgrass dataset due to the larger number of samples available in the linear range. Figure \ref{fig:7}(a) presents the resulting distributions at three sample wind speeds following application of a first order low-pass filter with a cutoff frequency of 1 Hz. Figure \ref{fig:7}(b) presents the peak of the distributions for all wind speeds. The distribution peak was selected to avoid potential bias from the tails. The difference in the average direction is less than $1^{\circ}$. Thus, VA is comparable to the sonic anemometer used in this study which has a directional resolution of $1^{\circ}$. Moreover, VA exhibits less variability, likely due to being derived from a 2D field average as opposed to the point-wise anemometer measurement.

\section{Conclusions}\label{sec:filetypes}
The present work has identified the sigmoidal structural response curve as a feature common to three species of vegetation with distinct morphology, size, and material properties. The striking collapse of response data in figure \ref{fig:5}(b) suggests that this framework can be used to conceptualize flow-structure interactions in a broader array of vegetation exposed to wind. The three-regime structure of the sigmoid, i.e. at low, intermediate, and high winds, can be understood in terms of vegetation response. At wind speeds below a threshold, vegetation motion shows limited sensitivity to increasing wind. In high winds, a similar insensitivity occurs, but due to distinct physical factors. Here, we observe that the vegetation becomes maximally deflected, with further increases in wind speed unable to achieve further deflection of the vegetation. A region of nearly linear correlation between wind speed and vegetation speed exists. However, quantitative application of VA in the regime requires a priori mapping of the tree kinematics to the corresponding region of the sigmoidal response curve. Hence, at least one point of site-specific calibration appears necessary in order to achieve  VA based on vegetation displacement statistics. Future work may endeavor to combine the present physics-based approach with data-driven methods that could infer the sigmoidal response mapping based on a library of other vegetation characteristics, such as the tree species, morphology, age, leaf cover, or other visible physical attributes.

\begin{Backmatter}

\paragraph{Acknowledgements}
The authors gratefully acknowledge M. Fu and N. Esparza-Duran for the assistance with the experiments; S. Anuszczyk, M. Cordeiro, M. Fu, P. Gunnarson, S. Madruga, N. Mohebbi, and N. Wei for watering the plants. 

\paragraph{Funding Statement}
This work was supported by the National Science Foundation (grant 2019712), Heliogen, and CAST at Caltech.

\paragraph{Declaration of Interests}
The authors report no conflict of interest.

\paragraph{Data Availability Statement}
The data used in this work will be made available through Mendelay Data.

\paragraph{Author Contributions}
Conceptualization: RHG \& JOD. Methodology: RHG \& JOD. Investigation: RHG. Software: RHG. Data analysis: RHG \& JOD. Funding acquisition: JOD \& RHG.

\paragraph{Supplementary Material}
Additional information can be found in the supplementary material.

\end{Backmatter}

\end{document}


\title[Physical constraints on visual anemometry]{Supplementary Material}

\author [Roni H. Goldshmid and John O. Dabiri]{Roni H. Goldshmid$^{1}$ {\href{https://orcid.org/0000-0001-9095-3259}{\includegraphics[width=0.02\textwidth]{orcid_logo}}}}

\author[Roni H. Goldshmid and John O. Dabiri]{John O. Dabiri$^{1^\ast,2}$ {\href{https://orcid.org/0000-0002-6722-9008}{\includegraphics[width=0.02\textwidth]{orcid_logo}}}}
 
\address[1]{Graduate Aerospace Laboratories, California Institute of Technology, Pasadena, CA 91125, USA}
\address[2]{Mechanical Engineering, California Institute of Technology, Pasadena, CA 91125, USA}

\corres{*}{Corresponding author. E-mail:
\emaillink{jodabiri@caltech.edu}}


\date{\textbf{Received:} XX 2022; \textbf{Revised:} XX XX 2022; \textbf{Accepted:} XX XX 2022}

\maketitle

\section{Weibull Distribution Fits}
\subsection{Weibull Fit Parameters}
Temporal statistics of the wind speed and vegetation displacements were fitted to Weibull distributions. The Weibull fit parameters and confidence intervals are presented in table S\ref{tab:fitVals}. 
    \begin{table}
      \begin{center}
      \begin{tabular}{lccccc}
          Vegetation    &        & $\lambda_w$ m/s         & $k_w$           & $\lambda_c$ m/s       & $k_c$ \\[3pt]
           Oak          & low    & $3.227\pm0.032$         & $04.16\pm0.13$  & $0.038\pm0.000$       & $02.09\pm0.03$\\
           Oak          & medium & $5.488\pm0.044$         & $04.78\pm0.14$  & $0.144\pm0.001$       & $02.58\pm0.04$\\
           Oak          & high   & $8.645\pm0.082$         & $05.48\pm0.21$  & $0.246\pm0.002$       & $03.25\pm0.06$\\
           Camphor      & low    & $3.173\pm0.020$         & $05.74\pm0.15$  & $0.066\pm0.001$       & $02.20\pm0.03$\\
           Camphor      & medium & $5.448\pm0.034$         & $04.92\pm0.11$  & $0.150\pm0.001$       & $02.70\pm0.03$\\
           Camphor      & high   & $8.472\pm0.117$         & $04.97\pm0.26$  & $0.182\pm0.002$       & $03.43\pm0.09$\\
           Bullgrass    & low    & $3.290\pm0.057$         & $03.00\pm0.12$  & $0.016\pm0.000$       & $06.14\pm0.11$\\
           Bullgrass    & medium & $5.562\pm0.084$         & $03.45\pm0.14$  & $0.036\pm0.000$       & $04.43\pm0.08$\\
           Bullgrass    & high   & $7.494\pm0.093$         & $04.18\pm0.16$  & $0.058\pm0.001$       & $06.86\pm0.32$\\
      \end{tabular}
      \caption{Weibull distribution parameters MLE of the wind (subscript $w$) and canopy displacement (subscript $c$) speeds measured. $\lambda$ is scale factor and $k$ is shape factor of the Weibull distribution. These correspond to figures 3 and 4 in the main text.}
      \label{tab:fitVals}
      \end{center}
    \end{table}
\clearpage

\subsection{Q-q Analysis}
Q-q curves of the wind speed and vegetation displacement distributions are presented below. These correspond to the oak tree (figure S\ref{fig:S1}), camphor tree (figure S\ref{fig:S2}), and bullgrass (figure S\ref{fig:S3}). The intermediate wind speed curves (middle column) are the same as figures 3(d)-(f) and 4(d)-(f) in the main text. The resulting linear correlations in the majority of the data (i.e., quartiles 1-3) suggest the sample data likely comes from the Weibull distribution. 

 \begin{figure}[!htb]
      \centerline{\includegraphics[width=.8\textwidth]{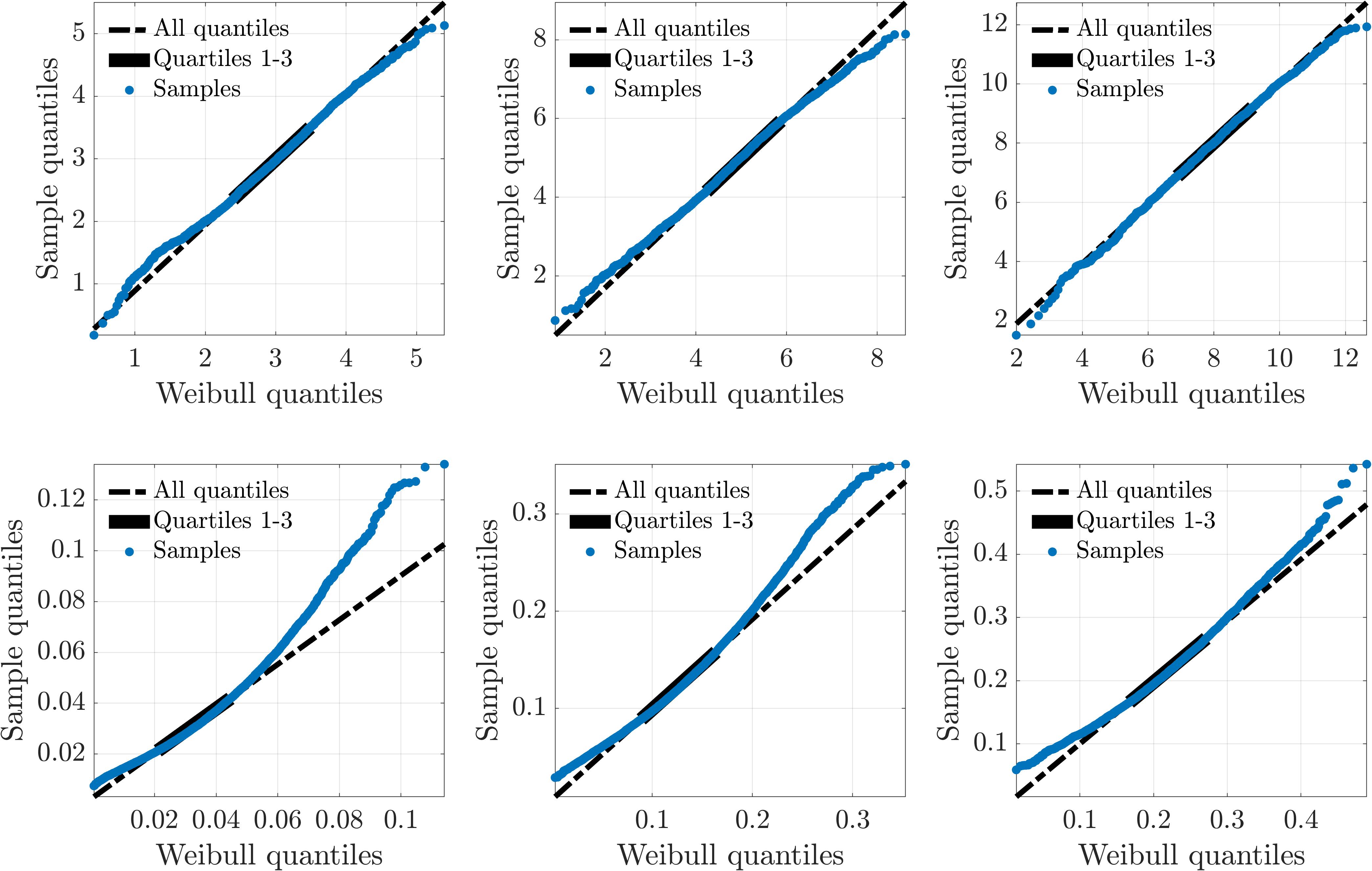}}
      \caption{(a)-(c) Example q-q curves of the wind speed distributions in the presence of the oak tree within the 3 m/s, 5 m/s, and 8 m/s bins, respectively. (d)-(f) Example q-q curves of the vegetation displacement distributions in the presence of the oak tree within the 3 m/s, 5 m/s, and 8 m/s bins, respectively. The solid reference line connects the first and third quartiles, corresponding to 75\% of the data, and the dashed reference line extends the solid line to the ends of the samples.}
    \label{fig:S1}
    \end{figure}
    
 \begin{figure}[!htb]
      \centerline{\includegraphics[width=.8\textwidth]{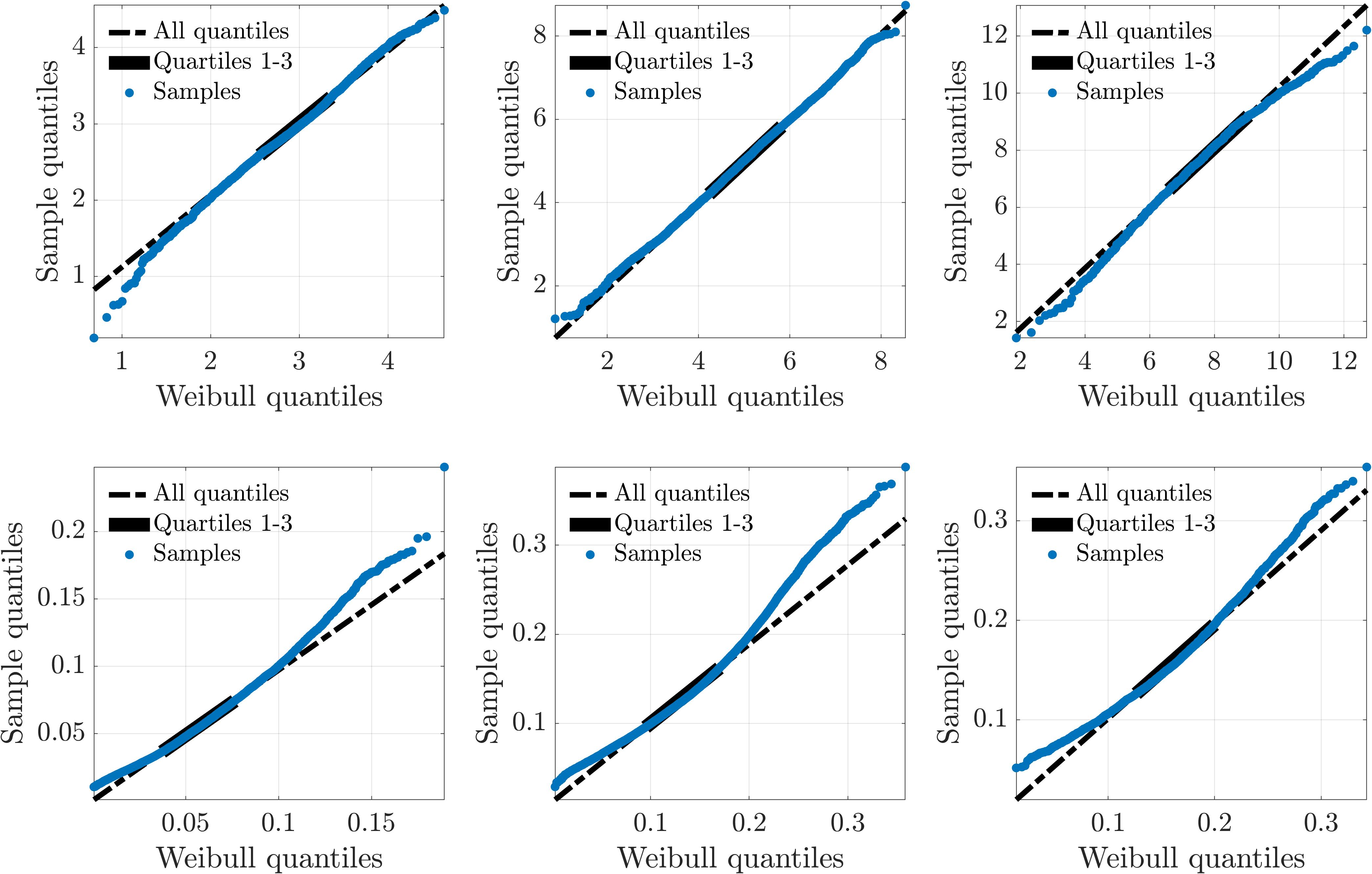}}
      \caption{(a)-(c) Example q-q curves of the wind speed distributions in the presence of the camphor tree within the 3 m/s, 5 m/s, and 8 m/s bins, respectively. (d)-(f) Example q-q curves of the vegetation displacement distributions in the presence of the camphor tree within the 3 m/s, 5 m/s, and 8 m/s bins, respectively. The solid reference line connects the first and third quartiles, corresponding to 75\% of the data, and the dashed reference line extends the solid line to the ends of the samples.}
    \label{fig:S2}
    \end{figure}
    
  \begin{figure}[!htb]
      \centerline{\includegraphics[width=.8\textwidth]{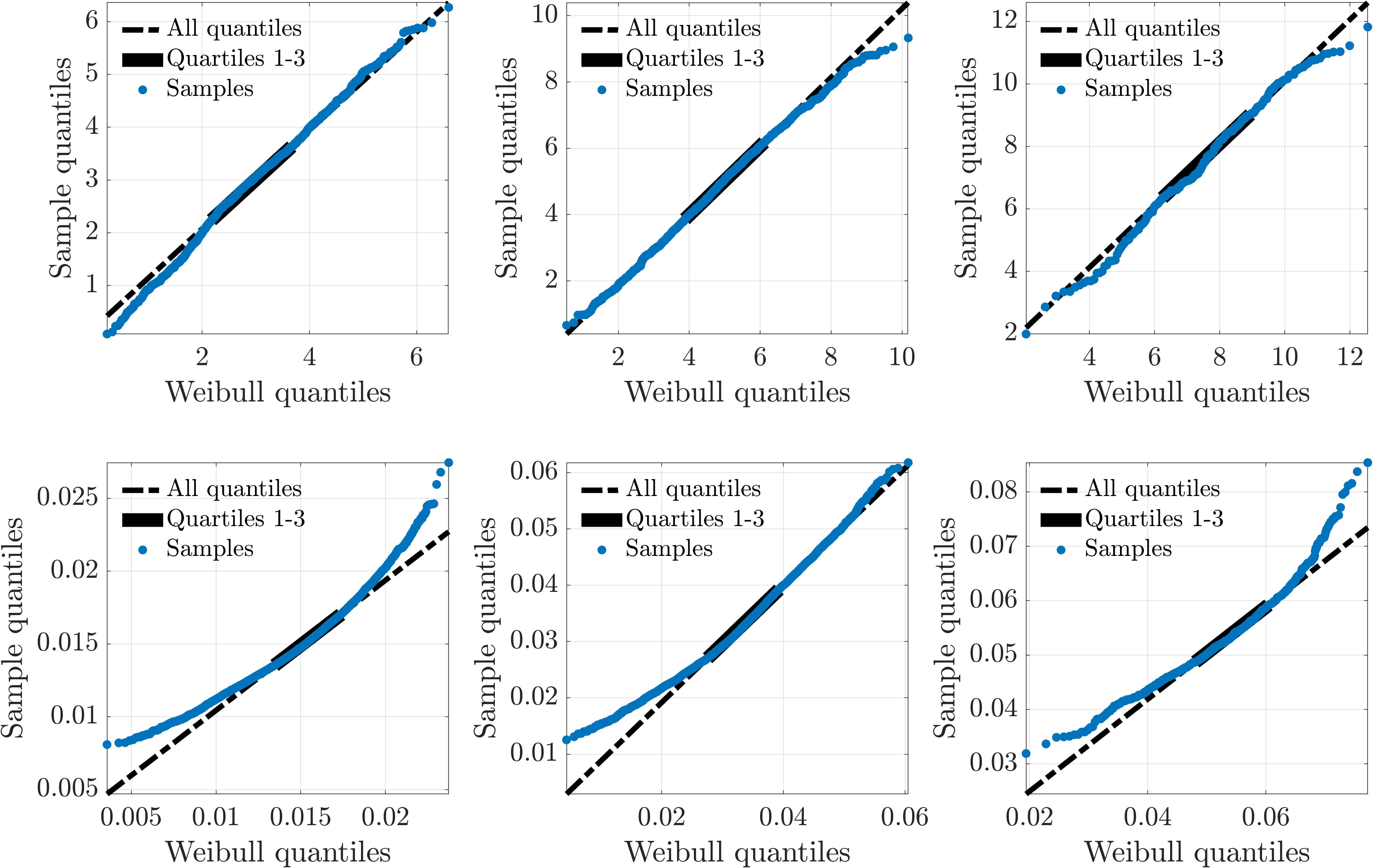}}
      \caption{(a)-(c) Example q-q curves of the wind speed distributions in the presence of the bullgrass within the 3 m/s, 5 m/s, and 8 m/s bins, respectively. (d)-(f) Example q-q curves of the vegetation displacement distributions in the presence of the bullgrass within the 3 m/s, 5 m/s, and 8 m/s bins, respectively. The solid reference line connects the first and third quartiles, corresponding to 75\% of the data, and the dashed reference line extends the solid line to the ends of the samples.}
    \label{fig:S3}
    \end{figure}
    
\clearpage
\section{Slenderness Relationship with Sigmoid Parameters}
\begin{figure}[!htb]
      \centerline{\includegraphics[width=0.6\textwidth]{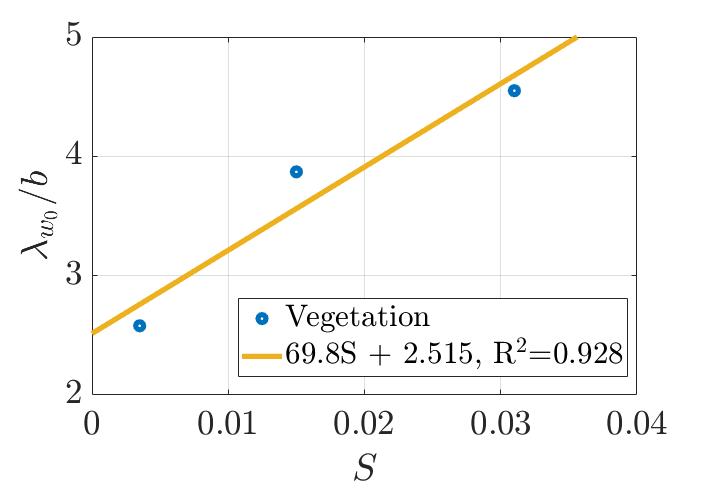}}
      \caption{Relationship between dimensionless ratio of $\lambda_{w_0}/b$ and the slenderness measure. The linear fit parameters are presented in the legend.}
    \label{fig:S4}
    \end{figure}

Figure S\ref{fig:S4} presents the relationship between dimensionless ratio of $\lambda_{w_0}/b$ and the slenderness estimate of the vegetation studied. Slenderness, $S$, estimates are taken from table 1 in the main text. Parameters $\lambda_{w_0}$ and $b$ are taken from table 2 in the main text.

